\documentclass[12pt]{iopart}

\usepackage{graphicx}
\usepackage{hyperref}
\usepackage{url}

\begin{document}

\title[]{Universal patterns of inequality}

\author{Anand Banerjee and Victor~M.~Yakovenko}

\address{Department of Physics, University of Maryland, College Park, Maryland 20742-4111, USA}

\begin{abstract}
Probability distributions of money, income, and energy consumption per capita are studied for ensembles of economic agents.  The principle of entropy maximization for partitioning of a limited resource gives exponential distributions for the investigated variables.  A non-equilibrium difference of money temperatures between different systems generates net fluxes of money and population.  To describe income distribution, a stochastic process with additive and multiplicative components is introduced.  The resultant distribution interpolates between exponential at the low end and power law at the high end, in agreement with the empirical data for USA.  We show that the increase of income inequality in USA originates primarily from the increase of the income fraction going to the upper tail, which now exceeds 20\% of the total income.  Analyzing the data from the World Resources Institute, we find that the distribution of energy consumption per capita around the world can be approximately described by the exponential function.  Comparing the data for 1990, 2000, and 2005, we discuss the effect of globalization on the inequality of energy consumption.
\end{abstract}

\pacs{
89.65.Gh, 
89.75.Da, 
05.20.Gg, 
88.05.Jk  
}

\submitto{New Journal of Physics}
\maketitle

\section{Introduction}

Two types of approaches are utilized in sciences to describe the natural world around us.  One approach is suitable for systems with a small number of degrees of freedom, such as a harmonic oscillator, a pair of gravitating bodies, and a hydrogen atom.  In this case, the goal is to formulate and solve dynamical equations of motion of the system, be it within Newtonian, relativistic, or quantum mechanics.  This approach is widely used beyond physics  to study dynamical systems in chemistry, biology, economics, etc.  In the opposite limiting case, we deal with systems consisting of a very large number of degrees of freedom.  In such cases, statistical description is employed, and the systems are characterized by probability distributions.  In principle, it should be possible to derive statistical description from microscopic dynamics, but it is rarely feasible in practice.   Thus, it is common to use general principles of the theory of probabilities to describe statistical systems, rather than to derive their properties from microscopic equations of motion.  Statistical systems are common in physics, chemistry, biology, economics, etc.

Any probability distribution can be thought of as representing some sort of ``inequality'' among the constituent objects of the system, in the sense that the objects have different values of a given variable.  Thus, a study of probability distributions is also a study of inequality developing in a system for statistical reasons.  To be specific, let us consider an economic system with a large number of interacting agents.  In the unrealistic case where all agents have exactly the same values of economic variables, the system can be treated as a single agent called the ``representative agent.''  This approach is common in traditional economics, but, by construction, it precludes a study of inequality among the agents.  However, social and economic inequality is ubiquitous in the real world, and its characterization and understanding are very important issues.

In this paper, we apply the well-developed methods of statistical physics to economics and society in order to gain insights into probability distributions and inequality in these systems.  We consider three specific cases: the distributions of money, income, and global energy consumption.  In all three cases, the common theme is entropy maximization for partitioning of a limited resource among multiple agents.  Despite the difference in the nature of the considered variables, we find a common pattern of inequality in these cases.   This approach can be also useful for studying other statistical systems beyond the three specific cases considered in this paper. 

Applications of these ideas to money and income have been published in the literature before: see review \cite{RMP}.  To introduce these ideas and to make the paper self-contained, we briefly review the applications to money and income in \sref{sec:money} and \sref{sec:income}.  \Sref{sec:income} also shows the latest available data for income distribution in 2007, not published before.  In \sref{sec:energy}, we present a quantitative study of the probability distribution of energy consumption per capita around the world.  This is a new kind of study that, to the best of our knowledge, has not appeared before in the literature.

\section{Statistical mechanics and thermodynamics of money}
\label{sec:money}

\subsection{Entropy maximization for division of a limited resource}
\label{sec:entropy}

Let us consider a general mathematical problem of partitioning (dividing) a limited resource among a large number of agents.  The solution of this problem is similar to the derivation of the Boltzmann-Gibbs distribution of energy in physics \cite{Wannier}.  To be specific, let us apply it to the probability distribution of money in a closed economic system.

Following \cite{AAD00}, let us consider a system consisting of $N$ economic agents.  At any moment of time, each agent $i$ has a money balance $m_i$.  Agents make pairwise economic transactions with each other.  As a result of a transaction, the money $\Delta m$ is transferred from an agent $i$ to an agent $j$, so their money balances change as follows
\begin{eqnarray}
  && m_i\;\rightarrow\; m_i'=m_i-\Delta m,
\nonumber \\
  && m_j\;\rightarrow\; m_j'=m_j+\Delta m.
\label{transfer}
\end{eqnarray}
The total money of the two agents before and after transaction remains the same
\begin{equation}
  m_i+m_j=m_i'+m_j',
\label{conservation}
\end{equation}
i.e.,\ there is a local conservation law for money.  It is implied that the agent $j$ delivers some goods or services to the agent $i$ in exchange for the money payment $\Delta m$.  However, we do not keep track of what is delivered and only keep track of money balances.  Goods, such as food, can be produced and consumed, so they are not conserved.

Rule (\ref{transfer}) for the transfer of money is analogous to the transfer of energy from one molecule to another in molecular collisions in a gas, and rule (\ref{conservation}) is analogous to conservation of energy in such collisions.  It is important to recognize that ordinary economic agents cannot ``manufacture'' money (even though they can produce and consume goods).  The agents can only receive money from and give it to other economic agents.  In a closed system, the local conservation law (\ref{conservation}) implies the global conservation law for the total money $M=\sum_im_i$ in the system.  In the real economy, $M$ may change due to money emission by the central government or central bank, but we will not consider these processes here.  Another possible complication is debt, which may be considered as negative money.  Here we consider a model where debt is not permitted, so all money balances are non-negative $m_i\geq0$.

After many transactions between different agents, we expect that a stationary probability distribution of money would develop in the system.  It can be characterized as follows.  Let us divide the money axis $m$ into the intervals (bins) of a small width $m_*$ and label them with an integer variable $k$.  Let $N_k$ be the number of agents with the money balances between $m_k$ and $m_k+m_*$.\footnote{Throughout the paper, we use the indices $k$ and $n$ to label the money bins $m_k$ and the indices $i$ and $j$ to label the individual money balances $m_i$ of the agents.}  Then, the probability to have a money balance in this interval is $P(m_k)=N_k/N$.  We would like to find the stationary probability distribution of money $P(m)$, which is achieved in statistical equilibrium.

Because the total money $M$ in the system is conserved, the problem reduces to partitioning (division) of the limited resource $M$ among $N$ agents.  One possibility is an equal division, where each agents gets the same share $M/N$ of the total money.  However, such an equal partition would be extremely improbable.  It is more reasonable to obtain  the probability distribution of money from the principle of entropy maximization.  Let us consider a certain set of occupation numbers $N_k$ of the money bins $m_k$.  The multiplicity $\Omega$ is the number of different realizations of this configuration, i.e., the number of different placements of the agents into the bins preserving the same set of occupation numbers $N_k$.  It is given by the combinatorial formula in terms of the factorials\footnote{Notice that human agents, unlike particles in quantum physics, are distinguishable.}
\begin{equation}
  \Omega=\frac{N!}{N_1!\,N_2!\,N_3!\,\ldots}.
\label{multiplicity}
\end{equation}
The logarithm of multiplicity is called the entropy $S=\ln\Omega$.  In the
limit of large numbers, we can use the Stirling approximation for the factorials
\begin{equation}
  S=N\ln N - \sum_k N_k\ln N_k =
  -\sum_k N_k\ln\left(\frac{N_k}{N}\right).
\label{S}
\end{equation}

In statistical equilibrium, the entropy $S$ is maximized with respect to the numbers $N_k$ under the constraints that the total number of agents $N=\sum_k N_k$ and the total money $M=\sum_k m_k N_k$ are fixed.  To solve this problem, we introduce the Lagrange multipliers $\alpha$ and $\beta$ and construct the modified entropy
\begin{equation}
  \tilde S = S + \alpha\sum_k N_k - \beta\sum_k m_k N_k.
\label{S'}
\end{equation}
Maximization of entropy is achieved by setting the derivatives $\partial\tilde S/\partial N_k$ to zero for each $N_k$.  Substituting \eref{S} into \eref{S'} and taking the derivatives\footnote{Notice that $N=\sum_k N_k$ in \eref{S} should be also differentiated with respect to $N_k$.}, we find that the equilibrium probability distribution of money $P(m)$ is an exponential function of $m$
\begin{equation}
  P(m_k)=\frac{N_k}{N}=e^{\alpha-\beta m_k}=e^{-(m_k-\mu)/T}.
\label{P(m)}
\end{equation}
Here the parameters $T=1/\beta$ and $\mu=\alpha/T$ are the analogs of temperature and chemical potential for money.  Their values are determined by the constraints
\begin{equation}
  1=\frac{\sum_k N_k}{N}=\sum_k P(m_k)=\int\limits_0^\infty \frac{dm}{m_*}\,e^{-(m-\mu)/T}  \quad \Rightarrow \quad \mu=-T\ln\left(\frac{T}{m_*}\right),
\label{mu}
\end{equation}
\begin{equation}
  \langle m\rangle=\frac{M}{N}=\frac{\sum_k m_k N_k}{N}=\sum_k m_k P(m_k)
  =\int\limits_0^\infty \frac{dm}{m_*}\,m\,e^{-(m-\mu)/T}=T.
\label{T}
\end{equation}
We see that the money temperature $T=\langle m\rangle$ \eref{T} is nothing but the average amount of money per agent.  The chemical potential $\mu$ \eref{mu} is a  decreasing function of $T$.

\Eref{P(m)} shows quite generally that division of a conserved limited resource using the principle of entropy maximization results in the exponential probability distribution of this resource among the agents.  In physics, the ``limited resource'' is the energy $E$ divided among $N$ molecules of a gas, and the result is the Boltzmann-Gibbs distribution of energy \cite{Wannier}.  The exponential distribution of money \eref{P(m)} was proposed in \cite{AAD00}, albeit without explicit discussion of the chemical potential, as well as in \cite{Mimkes-2000}.  Various models for kinetic exchange of money are reviewed in \cite{RMP} and in the popular article \cite{PhysNews}.  The applicability of the underlying assumptions of money conservation and random exchange of money is discussed in \cite{RMP} and \cite{CurrentScience}.  The analogy between energy and money is mentioned in some physics textbooks \cite{Schroeder}, but not developed in detail.

\subsection{Flow of money and people between two countries with different temperatures}

To illustrate some consequences of the statistical mechanics of money, let us consider two systems with different money temperatures $T_1>T_2$.  These can be two countries with different average amounts of money per capita: the ``rich'' country with $T_1$ and the ``poor'' with $T_2$.\footnote{For simplicity, let us assume that the two countries use the same currency, or the currency exchange rate is fixed, so that an equivalent currency can be used.}  Suppose a limited flow of money and agents is permitted between the two systems.  Given that the variation $\delta \tilde S$ vanishes due to maximization under constraints, we conclude from \eref{S'} that 
\begin{equation}
  \delta S = \beta\,\delta M - \alpha\,\delta N \quad \Leftrightarrow \quad
  \delta M = T\,\delta S + \mu\,\delta N.
\label{dS}
\end{equation}
If $\delta M$ and $\delta N$ denote the flow of money and agents from system 1 to system 2, then the change of the total entropy of the two systems is
\begin{equation}
  \delta S = (\beta_2-\beta_1)\,\delta M -(\alpha_2-\alpha_1)\,\delta N
  = \left(\frac{1}{T_2}-\frac{1}{T_1}\right)\delta M
  + \ln\left(\frac{T_2}{T_1}\right)\delta N.
\label{S12}
\end{equation}
According to the second law of thermodynamics, the total entropy should be increasing, so $\delta S\geq0$.  Then, the first term in \eref{S12} shows that money should be flowing from the high-temperature system (rich country) to the low-temperature system (poor country).  This is called the trade deficit -- a systematic net flow of money from one country to another, which is best exemplified by the trade between USA and China.  The second term in \eref{S12} shows that the agents would be flowing from high to low chemical potential, which corresponds to immigration from a poor to a rich country.  Both trade deficit and immigration are widespread global phenomena.  The direction of these processes can be also understood from \eref{T}.  The two systems are trying to equilibrate their money temperatures $T=M/N$, which can be achieved either by changing the numerators due to money flow or the denominators due to people flow.

\subsection{Thermodynamics of money and wealth}
\label{sec:Thermodynamics}

Thermal physics has two counterparts: statistical mechanics and thermodynamics.  Statistical mechanics of money was outlined in \sref{sec:entropy}.  Is it possible to construct an analog of thermodynamics for money?  Many attempts were made in the literature, but none was completely successful: see reviews \cite{PhysNews} and \cite{Mirowski-book}.

One of the important concepts in thermodynamics is the distinction between heat and work.  In statistical physics, this distinction can be microscopically interpreted as follows \cite{Wannier}.  The internal energy of the system is $U=\sum_k \varepsilon_k N_k$, where $\varepsilon_k$ is an energy level, and $N_k$ is the occupation number of this level.  Suppose the energy levels $\varepsilon_k(\lambda)$ depend on some external parameters $\lambda$, such as the volume of a box in quantum mechanics, an external magnetic field acting on spins, etc.  Then, the variation of $U$ contains two terms $\delta U=\sum_k(\delta\varepsilon_k) N_k+\sum_k\varepsilon_k(\delta N_k)=\delta W + \delta Q$.  The first term has mechanical origin and comes from the variation $\delta\varepsilon_k=(\partial\varepsilon_k/\partial\lambda)\,\delta\lambda$ of the energy levels due to changes of the external parameters $\lambda$.  This term is interpreted as the work $\delta W$ done on the system externally.  The second term has statistical origin and comes from the changes $\delta N_k$ in the occupation numbers of the energy levels.  This term is interpreted as the heat $\delta Q$.

An analog of this construction does not seem to exist for money $M=\sum_k m_k N_k$.  A variation $\delta M=\sum_k m_k (\delta N_k)$ is possible due to changes in the occupation numbers, but there is no analog of the variation $\delta m_k$ of the ``money levels'' due to changes in some external parameters.  Thus, we can only define the heat term, but not the work term in the money variation.  Indeed, \eref{dS} is the analog of the first law of thermodynamics for money, but there is no term corresponding to work in this equation.

Nevertheless, statistical mechanics of money can be extended to a form somewhat resembling conventional thermodynamics, if we take into account the material property of the agents.  Let us define the wealth $w_i$ of an agent $i$ as a sum of two terms.  One term represents the money balance $m_i$, and another term the material property, such as a house, a car, stocks, etc.  For simplicity, let us consider only one type of property, so that the agent has $v_i$ physical units of this property.  In order to determine the monetary value of this property, we need to know the price $P$ per unit.  Then, the wealth of the agent is $w_i=m_i+Pv_i$.  Correspondingly, the total wealth $W$ in the system is\footnote{From now on, we use the letter $W$ to denote wealth, not work.}
\begin{equation}
  W = M + PV,
\label{W}
\end{equation}
where $V=\sum_i v_i$ is the total ``volume'' of the property in the system.  If money $M$ is analogous to the internal energy $U$ in statistical physics, then wealth $W$ is analogous to the enthalpy $H$.  The wealth $W$ includes not only the money $M$, but also the money equivalent necessary to acquire the volume $V$ of property at the price $P$ per unit.

Let us consider the differential of wealth
\begin{equation}
  dW = dM + P\,dV + V\, dP = V\, dP.
\label{dW}
\end{equation}
Here the first two terms cancel out, and only the last term remains.  Indeed, when the volume $dV>0$ of property is acquired, the money $dM=-P\,dV<0$ is paid for the property, i.e.,\ money is exchanged for property.  \Eref{dW} is also valid at the level of individual agents, $dw_i=v_i\,dP$.  These equations show that wealth changes only when the price $P$ changes.

\begin{figure}
\centering
\includegraphics[width=0.5\linewidth]{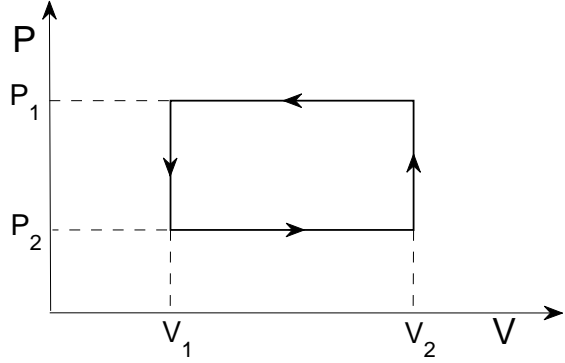}
\caption{A closed cycle of speculation or trading.  $V$ and $P$ represents the volume and price of goods.}
\label{fig:PV}
\end{figure}

To advance the analogy with thermodynamics, let us consider a closed cycle in the $(V,P)$ plane illustrated in \fref{fig:PV}.  This cycle can be interpreted as a model of stock market speculation, in which case $V$ is the volume of stock held by a speculator.  Starting from the lower left corner, the speculator purchases the stock at the low price $P_2$ and increases the owned volume from $V_1$ to $V_2$.  Then, the price increases from $P_2$ to $P_1$.  At this point, the speculator sells the stock at the high price $P_1$, reducing the owned volume from $V_2$ to $V_1$.  Then, the price of the stock drops to the level $P_2$, and the cycle can be repeated.  From \eref{dW}, we find that the wealth change of the speculator is $\Delta W=\oint V\,dP$, which is the area $(P_1-P_2)(V_2-V_1)$ enclosed by the cycle in \fref{fig:PV}.  From \eref{W}, we also find that $\Delta W=\Delta M$, because $P$ and $V$ return to the initial values at the end of the cycle.  Thus, the monetary profit $\Delta M$ is given by the area enclosed by the cycle.  This money is extracted by the speculator from the other players in the market, so the conservation law of money is not violated.  In the ideal economic equilibrium, there should be no price changes allowing one to make systematic profits, which is known as the ``no-arbitrage theorem''.  However, in the real market, significant rises and falls of stock prices do happen, especially during speculative bubbles.

The cycle in \fref{fig:PV} also illustrates the trade between China and USA.  Suppose a trade company pays money $M_2$ to buy the volume $V_2-V_1$ of the products manufactured in China at the low price $P_2$.  After shipping across Pacific Ocean, the products are sold in USA at the high price $P_1$, and the company receives money $M_1$.  Empty ships return to China, and the cycle repeats.  As shown in  \cite{AAD00}, the price level $P$ is generally proportional to the money temperature $T$.  Thus, the profit rate in this cycle is
\begin{equation}
  \mbox{profit rate} = \frac{M_1-M_2}{M_2} = \frac{P_1-P_2}{P_2} 
  = \frac{T_1-T_2}{T_2}.
\label{profit}
\end{equation}
By analogy with physics, one can prove that \eref{profit} gives the highest possible profit rate for the given temperatures $T_1$ and $T_2$.  Indeed, from \eref{dS} with $\delta N=0$, we find that $M_1=T_1\,\Delta S_1$ and $M_2=T_2\,\Delta S_2$.  Under the most ideal circumstances, the total entropy of the whole system remains constant, so $\Delta S_1=\Delta S_2$.  Then, $M_1/M_2=T_1/T_2$, and \eref{profit} follows.  Here we assumed that the profit money $M_1-M_2$ has low (ideally zero) entropy, because this money is concentrated in the hands of just one agent or trading company and is not dispersed among many agents of the systems.

Thermal machines have cycles analogous to \fref{fig:PV}, and equation \eref{profit} is similar to the Carnot formula for the highest possible efficiency \cite{Wannier,Schroeder}.  The China-USA trade cycle resembles an internal combustion engine, where the purchase of goods from China mimics fuel intake, and the sales of goods in USA mimics expulsion of exhaust.  The net result is that goods are manufactured in China and consumed in USA.  The analogy between trade cycles and thermal machines was highlighted by Mimkes \cite{Mimkes-2005,Mimkes-2010}.  Although somewhat similar to \cite{Mimkes-2010}, our presentation emphasizes conceptual distinction between money and wealth and explicitly connects statistical mechanics and thermodynamics.

Empirical data on the international trade network between different countries were analyzed in several papers.  Serrano \textit{et al.}\ \cite{Vespignani} analyzed trade imbalances, defined as the \textit{difference} between exports and imports from one country to another.  The paper classified countries as net consumers and net producers of goods.  The typical examples are USA and China, respectively, as illustrated in Figure 2 of \cite{Vespignani} for 2000, in qualitative agreement with our discussion above.  In contrast, Bhattacharya \textit{et al.}\ \cite{ITN} studied trade volumes, defined as the \textit{sum} of exports and imports from one country to another.  The paper found that the trade volume $s$ of a country is proportional to the gross domestic product (GDP) of the country: $s\propto({\rm GDP})^\gamma$ with the exponent $\gamma\approx1$.  It means that the trade volume and GDP are \emph{extensive} variables in the language of thermodynamics, so the biggest volumes of trade are between the countries with the biggest GDPs.  In thermal equilibrium, money flows between two countries in both directions as payment for traded goods, but the money fluxes in the opposite directions are equal, so there is trade volume, but no trade imbalance.  Trade imbalance may develop when the two systems have different values of \emph{intensive} parameters, such as the money temperature.  Then, the direction of net money flow is determined by the sign of the temperature difference.

Of course, there may be other reasons and mechanisms for trade imbalance besides the temperature difference.  Normally, the flow of money from the high- to low-temperature system should reduce the temperature difference and eventually bring the systems to equilibrium.  Indeed, in the global trade, many formerly low-temperature countries have increased their temperatures as a result of such trade.  However, the situation with China is special, because the Chinese government redirects the flow of dollars back to USA by buying treasury bills from the US government.  As a result, the temperature difference remains approximately constant and does not show signs of equilibration.  The net result is that China supplies vast amounts of products to USA in exchange for debt obligations from the US government.  The long-term global consequences of this process remain to be seen.

\subsection{The circuit of money and the circuit of goods}
\label{sec:Circuits}

\Sref{sec:Thermodynamics} illustrates that there are two circuits in a well-developed market economy \cite{Mimkes-2010}.  One is the circuit of money, which consists of money payments between the agents for goods and services.  As argued in \sref{sec:entropy}, money is conserved in these transactions and, thus, can be modeled as flow of liquid, e.g., blood in the vascular system. 
(A hydraulic device, the MONIAC, was actually used by William Philips, the inventor of the famous Philips curve, to illustrate money flow in the economy \cite{PhysNews}.)  The second circuit is the flow of goods and services between the agents.  This circuit involves manufacturing, distribution, and consumption.  The goods and services are inherently not conserved.  They represent the material (physical) side of the economy and, arguably, are the ultimate goal for the well-being of a society.  In contrast, money represents the informational, virtual side of the economy, because money cannot be physically consumed.  Nevertheless, money does play a very important role in the economy by enabling its efficient functioning and by guiding resource allocation in a society.\footnote{Here we consider the modern fiat money, declared to be money by the central bank or government.  We do not touch the origin of money in the early history as some kind of special goods.}  The two circuits interact with each other when goods and services are traded (exchanged) for money.  However, money cannot be physically transformed into goods and vice versa.  To illustrate this point, we draw an analogy with fermions and bosons in physics.  While the ``circuits'' of fermions and bosons interact and transfer energy between each other, it is not possible to convert a fermion into a boson and vice versa.

The important consequence of this consideration is that an increase of material production in the circuit of goods and services does not have any direct effect on the amount of money in the monetary circuit.  The amount of money in the system depends primarily on the monetary policy of the central bank or government, who have the monopoly on issuing money.  Technological progress in material production does not produce any automatic increase of money in the system.  Thus, the expectation of continuous monetary growth, where the agents would be getting more and more money as a result of technological progress, is false.  It is not possible for all businesses to operate with profit on average, i.e., to have the greater total amount of money at the end of a cycle than at the beginning.  The agents can get more money on average only if the government decides to print money, i.e.,\ to increase the money temperature $T=M/N$.\footnote{For discussion of the issues related to debt, see the review \cite{RMP}.}  Thus, monetary growth of the economy is directly related to the deficit spending by the central government.  On the other hand, it is very well possible to have technological progress and an increase in the physical standards of living without monetary growth.  The monetary and physical circuits of the economy interact with each other, but they are separate circuits.  Unfortunately, this distinction is often blurred in the econophysics and economics literature \cite{Mimkes-2010}, as well as in the public perception.

\section{Two-class structure of income distribution}
\label{sec:income}

\subsection{Introduction}

Although the exponential probability distribution of money \eref{P(m)} was proposed 10 years ago \cite{AAD00,Mimkes-2000}, no direct statistical data on money distribution are available to verify this conjecture.  Normally, people do not report their money balances to statistical agencies.  Given that most people keep their money in banks, the distribution of balances on bank accounts can give a reasonable approximation of the probability distribution of money.  However, these data are privately held by banks and not available publicly.

On the other hand, a lot of statistical data are available on income distribution, because people report income to the government tax agencies.  To some extent, income distribution can also be viewed as a problem of partitioning of a limited resource, in this case of the total  annual budget.  Following \sref{sec:entropy}, we expect to find the exponential distribution for income.  Dr\u{a}gulescu and Yakovenko \cite{AAD01a} studied the data on income distribution in USA from the Internal Revenue Service (IRS) and from the US Census Bureau.  They found that income distribution is indeed exponential for incomes below 120 k\$ per year.  However, in the subsequent papers \cite{AAD01b,AAD03}, they also found that the upper tail of income distribution follows a power law, as was first pointed out by Pareto \cite{Pareto-book}.  So, the data analysis of income distribution in USA reveals coexistence of two social classes.  The lower class (about 97\% of population) is characterized by the exponential Boltzmann-Gibbs distribution, and the upper class (the top 3\% percent of the population) has the power-law Pareto distribution.  Time evolution of the income classes in 1983--2001 was studied by Silva and Yakovenko \cite{Silva}.  They found that the exponential distribution in the lower class is very stable in time, whereas the power-law distribution of the upper class is highly dynamical and volatile.  They concluded that the lower class is in thermal equilibrium, whereas the upper class is out of equilibrium.  

Many other papers investigated income distributions in different countries:  see the review \cite{RMP} for references.  The coexistence of two classes appears to be a universal feature of income distribution.  In this section, we present a unified description of the two classes within a single mathematical model.

\subsection{Income dynamics as a combination of additive and multiplicative stochastic processes}

The two-class structure of income distribution can be rationalized on the basis of a kinetic approach.  Suppose the income $r$ of an agent behaves like a stochastic variable.  Let $P(r,t)$ denote the probability distribution of $r$ at time $t$.  Let us consider a diffusion model, where the income $r$ changes by $\Delta r$ over a time period $\Delta t$.
Then, the temporal evolution of $P(r,t)$ is described by the Fokker-Planck
equation \cite{Kinetics}
  \begin{equation}
  \frac{\partial P(r,t)}{\partial t}=\frac{\partial}{\partial r}
  \left[A(r)P(r,t)\right]+ \frac{\partial^2}{\partial r^2}\left
  [B(r)P(r,t)\right].
  \label{diffusion}
  \end{equation}
The coefficients $A(r)$ and $B(r)$ are the drift and the diffusion terms, which are determined by the first and second moments of the income changes $\Delta r$ per unit time
  \begin{equation}
  A(r)  = -{\langle\Delta r\rangle \over \Delta t}, \quad
  B(r)  = {\langle(\Delta r)^2\rangle \over 2\Delta t}.
  \label{AB}
  \end{equation}
The stationary solution $P_{\rm s}(r)$ of \eref{diffusion} satisfies $\partial_tP_{\rm s}=0$; thus we obtain
  \begin{equation}
  \frac{\partial(BP_{\rm s})}{\partial r}=-AP_{\rm s}.
  \label{dP/dr}
  \end{equation}
The general solution of \eref{dP/dr} is
  \begin{equation}
  P_{\rm s}(r)=\frac{c}{B(r)}\exp\left(-\int^r\frac{A(r')}{B(r')}dr'\right),
  \label{stationary}
  \end{equation}
where $c$ is a normalization factor, such that $\int_0^{\infty}P_{\rm s}(r)\,dr =1$. 

In the lower class, the income comes from wages and salaries, so it is reasonable to assume that income changes are independent of income itself, i.e.,\ $\Delta r$ is independent of $r$.  This process is called the additive diffusion \cite{Silva}.  In this case, the coefficients in (\ref{diffusion}) are some constants $A_0$ and $B_0$. Then (\ref{stationary}) gives the exponential distribution
  \begin{equation}
  P_{\rm s}(r) = \frac{1}{T}\,e^{-r/T}, \qquad T=\frac{B_0}{A_0}.
  \label{exponential}
  \end{equation}
On the other hand, the upper-class income comes from bonuses, investments, and capital gains, which are calculated in percentages.  Therefore, for the upper class, it is reasonable to expect that $\Delta r\propto r$, i.e.,\ income changes are proportional to income itself.  This is known as the proportionality principle of Gibrat \cite{Gibrat-1931}, and the process is called the multiplicative diffusion \cite{Silva}.  In this case, $A=ar$ and $B=br^2$, and (\ref{stationary}) gives a power-law distribution
  \begin{equation}
  P_{\rm s}(r) \propto \frac{1}{r^{1+\alpha}}, \qquad \alpha=1+\frac ab.
  \label{power-law}
  \end{equation}
The multiplicative hypothesis for the upper class income was quantitatively verified in \cite{Aoki-2003} for Japan, where tax identification data are officially published for the top taxpayers.

The additive and multiplicative processes may coexist.  For example, an employee may receive a cost-of-living raise calculated in percentages (the multiplicative process) and a merit raise calculated in dollars (the additive process).  Assuming that these processes are uncorrelated, we find that $A=A_0+ar$ and $B=B_0+br^2=b(r_0^2+r^2)$, where $r_0^2=B_0/b$.  Substituting these expressions into (\ref{stationary}), we find
  \begin{equation}
  P_{\rm s}(r)=c\,\frac{e^{-(r_0/T)\arctan(r/r_0)}}
  {[1+(r/r_0)^2]^{1+a/2b}} .
  \label{arctan}
  \end{equation}
The distribution (\ref{arctan}) interpolates between the exponential law for low $r$ and the power law for high $r$, because either the additive or the multiplicative process dominates in the corresponding limit.  A crossover between the two regimes takes place at $r\sim r_0$, where the additive and multiplicative contributions to $B$ are equal.  The distribution (\ref{arctan}) has three parameters: the temperature $T=A_0/B_0$, the Pareto exponent $\alpha=1+a/b$, and the crossover income $r_0$.  It is a minimal model that captures the salient features of the two-class income distribution.  A formula similar to \eref{arctan} was also derived by Fiaschi and Marsili \cite{Fiaschi-Marsili} for a microscopic economic model, which is effectively described by \eref{diffusion}.

\subsection{Comparison with the personal income data from IRS}

\begin{table}
\centering
\begin{tabular}{|c|c|c|c|c|c|c|}
\hline
Year   & $T$ (k\$) & $\alpha$ & $r_0$ (k\$)  &  $r_*$ (k\$) &  $f$ (\%) &   $G$  \\ \hline
1996   &     33     &   1.63   &      76       &     116       &  11.8      &   0.55  \\
1997   &     35     &   1.57   &      79       &     120       &  13.8      &   0.56  \\
1998   &     36     &   1.55   &      80       &     122       &  16.4      &   0.57  \\
1999   &     38     &   1.54   &      83       &     124       &  16.8      &   0.58  \\
2000   &     40     &   1.34   &      105      &     150       &  18.4      &   0.59  \\
2001   &     41     &   1.46   &      99       &     152       &  14.4      &   0.56  \\
2002   &     41     &   1.51   &      99       &     154       &  12.6      &   0.55  \\
2003   &     41     &   1.48   &      101      &     156       &  13.7      &   0.56  \\
2004   &     43     &   1.41   &      105      &     158       &  16.7      &   0.58  \\
2005   &     44     &   1.36   &      108      &     159       &  19.5      &   0.59  \\
2006   &     46     &   1.36   &      107      &     160       &  20.5      &   0.60  \\
2007   &     48     &   1.34   &      113      &     166       &  21.5      &   0.60  \\	\hline
\end{tabular}
\caption{\label{table:IRS} $T$, $\alpha$, and $r_0$ are the parameters in \eref{arctan}, obtained by fitting the annual income data from IRS.  $r_*$ is the income separating the upper and lower classes.  $f$ is the fraction of income going to the upper class, given by \eref{f}.  $G$ is the Gini coefficient.}
\end{table}

In this section, we compare (\ref{arctan}) with the annual income data from IRS for the years 1996--2007 \cite{IRS-data}.  Because IRS releases the data with a delay of a couple of years, 2007 is the latest year for which the data are currently available.  The IRS data are given for a set of discrete income levels.  Thus, it is more practical to construct the cumulative distribution function (CDF), which is the integral $C(r)= \int_r^\infty P(r')\,dr'$ of the probability density.  For the probability density \eref{arctan}, $C(r)$ is not available in analytical form, therefore it has to be calculated by integrating $P_{\rm s}(r)$ numerically.  We use the theoretical CDF $C_{\rm t}(r)$ to fit the empirical CDF $C_{\rm e}(r)$ calculated from the IRS data.

Determining the best values of the three fitting parameters in the theoretical CDF is a computationally challenging task.  Thus, we do it step by step.  For each year, we first determine the values of $T$ and $\alpha$ by fitting the low-income part of $C_{\rm e}(r)$ with an exponential function and the high-income part with a power law.  Then, keeping these two parameters fixed, we determine the best value of $r_0$  by minimizing the mean-square deviation $\Sigma_n\ln^2[C_{\rm t}(r_n)/C_{\rm e}(r_n)]$ between the theoretical and empirical functions, where the sum is taken over all income levels  $r_n$ for which empirical data are available.  

\begin{figure}
\centering
\includegraphics[width=0.45\linewidth]{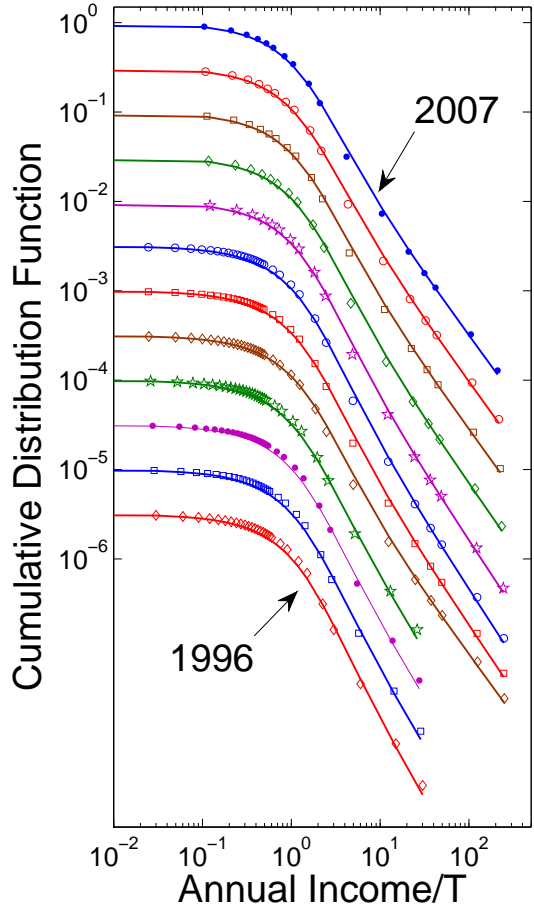}
\caption{Cumulative distribution functions constructed from the IRS data (symbols) and their fits with the theoretical distribution \eref{arctan}, shown in the log-log scale versus the normalized annual income $r/T$.  Plots for different years are shifted vertically for clarity.}
\label{fig:irsloglog}
\end{figure}

\Tref{table:IRS} shows the values of the fit parameters obtained for different years.  The data points for the empirical CDF and their fits with the theoretical CDF are shown in \fref{fig:irsloglog} in the log-log scale versus the normalized annual income $r/T$.  For clarity, the curves are shifted vertically for successive years. Clearly, the theoretical curves agree well with the empirical data, so the minimal model \eref{arctan} indeed captures the salient features of income distribution in USA.

In previous papers \cite{AAD03,Silva}, fits of the income distribution data were made only to the exponential \eref{exponential} and power-law \eref{power-law} functions.  The income $r_*$, where the two fits intersect, can be considered as a boundary between the two classes.  The values of $r_*$ are shown in \tref{table:IRS}.  We observe that the boundary $r_*$ between the upper and lower classes is approximately 3.5 times greater than the temperature $T$.  Given that the CDF of the lower class is exponential, we find that the upper class population is approximately $\exp(-r_*/T)=\exp(-3.5)=3\%$, which indeed agrees with our observations.

\subsection{The fraction of income in the upper tail and speculative bubbles}
\label{sec:f}

Let us examine the power-law tail in more detail.  Although the tail contains a small fraction of population, it accounts for a significant fraction $f$ of the total income in the system.  The upper-tail income fraction can be calculated as
  \begin{equation}
  f = \frac{R-N_{\rm e}T}{R} \approx \frac{R-NT}{R} =
  1- \frac{T}{\langle r\rangle}.
  \label{f}
  \end{equation}
Here $R$ is the total income, $N$ is the total number of people, and $\langle r\rangle = R/N$ is the average income for the whole system.  In addition, $N_{\rm e}$ is the number of people in the exponential part of the distribution, and $T$ is the average income of these people.  Since the fraction of people in the upper tail is very small, we use the approximation $N_{\rm e}\approx N$ in deriving the formula \eref{f} for $f$.  The values of $f$ deduced from the IRS data using \eref{f} are given in \tref{table:IRS}.

\begin{figure}
\centering
\includegraphics[width=0.7\linewidth]{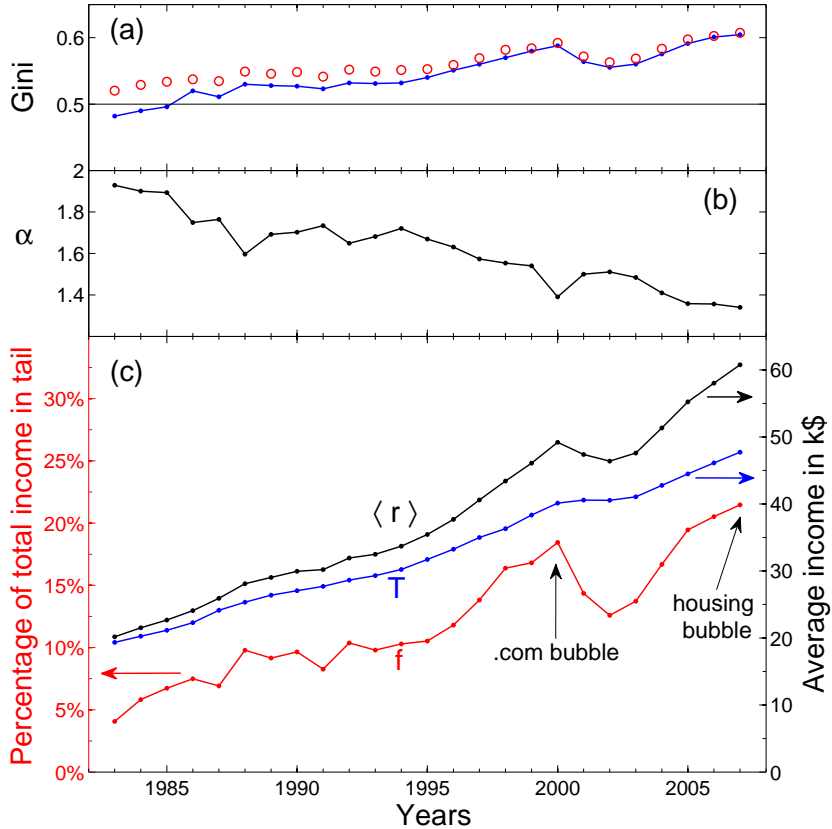}
\caption{{\bf (a)} Gini coefficient $G$ for the income distribution in USA in 1983--2007 (connected line) and the theoretical formula $G=(1+f)/2$ (open circles).  {\bf (b)} The exponent $\alpha$ of the power-law tail \eref{power-law} for income distribution.  {\bf (c)} The average income $\langle r\rangle$ in the whole system, the average income $T$ in the lower class (the temperature of the exponential part), and the percentage of income $f$ going to the upper tail.}
\label{fig:irsparam}
\end{figure}

Panel (c) in \fref{fig:irsparam} shows historical evolution of $\langle r\rangle$, $T$, and $f$ for the period 1983--2007.  We see that the average income $T$ of the lower class increases steadily without any large jumps.  In contrast, the fraction $f$ going to the upper class shows large variations and now exceeds 20\% of the total income in the system.  The maxima of $f$ are achieved at the peaks of speculative bubbles, first at the end of the ``.com'' bubble in 2000 and then at the end of the subprime mortgage bubble in 2007.  After the bubbles collapse, the fraction $f$ drops precipitously.  We conclude that the upper tail is highly dynamical and out of equilibrium.  The tail swells considerably during the bubbles, whereas the effect of the bubbles on the lower class is only moderate.  As a result, income inequality increases during bubbles and decreases when the bubbles collapse.

In the view of the argument about conservation of money presented in \sref{sec:entropy}, what is the source of money for the enormous increase of the upper tail income during speculative  bubbles?  The stock market bubble in the late 1990s was actually predicted in the book \cite{Dent-1993} published in 1993.  The prediction was based on the population data showing that the demographic wave of aging baby boomers will be massively investing their retirement money in the stock market in the second half of 1990s, which indeed happened.  The stock prices rose when millions of boomers paid for the stocks of ``.com'' companies.  When the demographic wave reached its peak around 2000 and the influx of money to the stock market started to saturate (at its highest level), the market crashed precipitously, and the population was left with worthless stocks.  One can see an analogy with the cycle in \fref{fig:PV}.  The net result of this bubble is the transfer of money from the lower to the upper class under the cover of ``retirement investment''.

As it is clear now, the second bubble in 2003--2007 was based on the enormous growth of debt due to proliferation of subprime mortgages.  As discussed in \cite{AAD00,RMP}, debt can be considered as negative money, because debt liabilities are counted with negative sign toward net worth of an individual.  The conservation law \eref{conservation} is still valid, but money balances $m_i$ can take negative values.  So, the first moment (the ``center of mass'') of the money distribution $\langle m\rangle=M/N$ remains constant.  However, now some agents can become super-rich with very high positive money balances at the expense of other agents plunging deeply into debt with negative money balances.  Thus, relaxing the boundary condition $m\geq0$ undermines stability of the Boltzmann-Gibbs distribution \eref{P(m)}.  This is what happened during the subprime mortgages bubble.  The money flowing to the upper tail were coming from the growth of the total debt in the system.  Eventually, the bubble collapsed when the debt reached a critical level.  Now the bailout effort by the government, effectively, represents the transfer of debt from economic agents to the government.  The overall result is that the income growth of the upper class in 2003--2007 was coming from the bailout money that the government is printing now.  As emphasized in \sref{sec:entropy} and \sref{sec:Circuits}, the government and central bank are the ultimate sources of new money because of the government monopoly on fiat money.

The discussion and the data presented in this section indicate that, by combining demographic data with the principle of money conservation, it may be possible to predict, to some degree, the macroeconomic behavior of the economy.  In fact, the book \cite{Dent-1993} predicted in 1993 that ``the next great depression will be from 2008 to 2023'' (page 16).  This is a stunning prediction 15 years in advance of the actual event.  For an update, see the follow-up book \cite{Dent-2009}.

\subsection{The power-law exponent of the upper tail}

Another parameter of the upper tail is the power-law exponent $\alpha$ in \eref{power-law}.  \Tref{table:IRS} and panel (b) in \fref{fig:irsparam} show historical evolution of $\alpha$ from 1983 to 2007.  We observe that $\alpha$ has decreased from about 2 to about 1.3.  The decrease of $\alpha$ means that the power-law tail is getting ``fatter'', i.e.,\  the inequality of income distribution increases.  It looks like the system is approaching dangerously closely to the critical value $\alpha=1$, where the total income in the tail $\int_{r_*}^\infty rP(r)\,dr$ would formally diverge \cite{PWA}.  On top of the gradual decrease, $\alpha$ dived down and up sharply around 1987 and 2000.  The dive-downs of $\alpha$ represent sharp increases of income inequality due to the bubbles, followed by crashes of the bubbles in 1987 and 2000 and subsequent contractions of the upper tail.  Thus, the behavior of the tail exponent $\alpha$ is qualitatively consistent with the behavior of the tail fraction $f$ discussed in \sref{sec:f}.  A similar behavior was found for Japan \cite{Aoki-2003}, where $\alpha$ jumped sharply from 1.8 to 2.1 between 1991 and 1992 due to  the crash of the Japanese market bubble.

During the times of bubbles, the sharp decrease of $\alpha$ is clearly a dynamical process, which cannot be described adequately by stationary equations.  On the other hand, during the time between bubbles, which economists may call ``recession'' or ``depression'', the market is quiet, and it may be possible to describe it using a stationary approach.  Even during these times, the power-law tail does not disappear, but the exponent $\alpha$ takes a relatively high value.  From the panel (b) in \fref{fig:irsparam}, it appears that the upper limit for $\alpha$ is about 2.  This limiting value is supported by other observations in the literature.  Analysis of Japanese data \cite{Aoki-2003} shows that $\alpha$ changes in the range between 1.8 to 2.2.   Dr\u{a}gulescu and Yakovenko \cite{AAD01b} found $\alpha=1.9$ for wealth distribution in UK for 1996.  Thus, we make a conjecture that $\alpha=2$ is a special value of the power-law exponent corresponding to a quiet, stationary market.

In order to understand what is special about $\alpha=2$, let us examine the moments of the income change $\Delta r$.  The first moment, $\langle\Delta r\rangle$ is always negative.  This condition ensures that $A>0$ in \eref{AB}, so that \eref{dP/dr} has a stationary solution.  The condition $\langle\Delta r\rangle<0$ indicates that, on average, everybody's income is decreasing due to the drift term, yet the whole income distribution remains stationary because of the diffusion term.  In stochastic calculus, the first $\langle\Delta r\rangle$ and the second $\langle(\Delta r)^2\rangle$ moments are of the same order in $\Delta t$, so they must be treated on equal footing.  Thus, instead of considering the changes in $r$, let us discuss how $r^2$ changes in time.  Using \eref{AB}, we find 
  \begin{equation}
  \langle\Delta(r^2)\rangle = \langle(r+\Delta r)^2-r^2\rangle 
  = 2r\langle\Delta r\rangle + \langle(\Delta r)^2\rangle 
  = 2(-rA + B)\,\Delta t.
  \label{Dr^2}
  \end{equation}
For the additive stochastic process \eref{exponential}, we find from \eref{Dr^2} that $\langle\Delta(r^2)\rangle>0$ for $r<T$ and $\langle\Delta(r^2)\rangle<0$ for $r>T$.  These conditions indicate a stabilizing tendency of the income-squares to move in the direction of the average income $T$.

Now, let us apply \eref{Dr^2} to the multiplicative process \eref{power-law}.  In this case, we find
  \begin{equation}
  \langle\Delta(r^2)\rangle = 2(-a+b)\,r^2\Delta t.
  \label{a=b}
  \end{equation}
For $a=b$, \eref{a=b} gives $\langle\Delta(r^2)\rangle=0$ for all $r$.  This condition can be taken as a criterion for the inherently stationary state of a power-law tail, because $r^2$ does not change (on average) for any $r$ in a scale-free manner.  From \eref{power-law}, we observe that the condition $a=b$ corresponds to the value $\alpha=1+a/b=2$, which is indeed the upper value of the power-law exponent observed for stationary, quiet markets:
  \begin{equation}
  \langle\Delta(r^2)\rangle=0 \quad \Leftrightarrow \quad a=b
  \quad \Leftrightarrow \quad \alpha=2 .
  \label{aplha=2}
  \end{equation}
On the other hand, for $a<b$, we find $\langle\Delta(r^2)\rangle>0$ and $\alpha<2$.  In this case, the income-square increases on average, which correlates with the upper tail expansion during the boom times.
Notice that the value $\alpha=2$ in \eref{aplha=2} is different from the value $\alpha=1$ found for the models of random saving propensity and earthquakes in \cite{PhysNews,CurrentScience,earthquake}.

\subsection{Lorenz plot and Gini coefficient for income inequality}

The standard way of representing income distribution in the economic literature is the Lorenz plot \cite{Kakwani-book}.  It is defined parametrically in terms of the two coordinates $x(r)$ and $y(r)$ depending on a parameter $r$
  \begin{equation}
  x(r) = \int\limits_0^{r}dr'P(r'), \quad
  y(r) = \frac{\int\limits_0^{r}dr'r'P(r')}{\int\limits_0^{\infty}dr'r'P(r')}.
  \label{Lorenz}
  \end{equation}
Here $x(r)$ is the fraction of the population with incomes below $r$, and $y(r)$ is the total income of this population, as a fraction of the total income in the system.  When $r$ changes from 0 to $\infty$, the variables $x$ and $y$ change from 0 to 1 producing the Lorenz plot in the $(x,y)$ plane.  The advantage of the Lorenz plot is that it emphasizes the data where most of the population is.  In contrast, the log-linear and log-log plots, like \fref{fig:irsloglog}, emphasize the upper tail, which corresponds to a small fraction of population, and where the data points are sparse.  Another advantage of the Lorenz plot is that all available data are represented within a finite area in the $(x,y)$ plane, whereas, in other plots, the upper end of the data at $r\to\infty$ is inevitably truncated.

\begin{figure}
\centering
\includegraphics[width=0.55\linewidth]{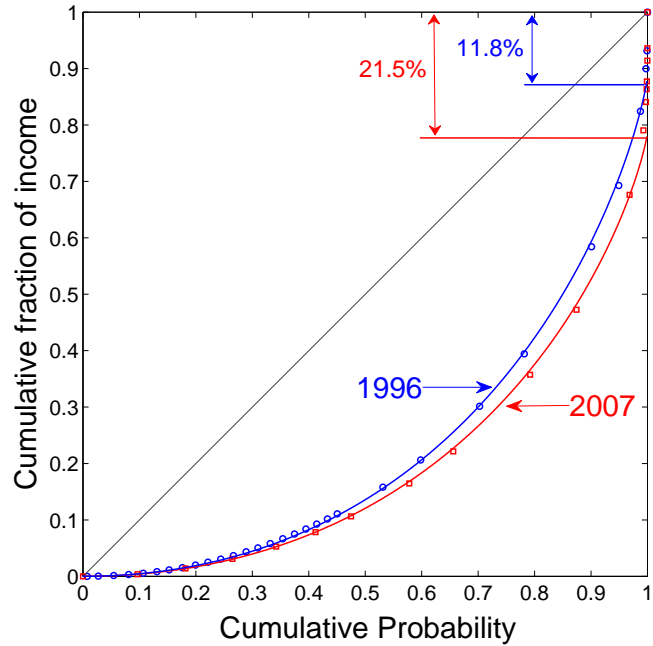}
\caption{Lorenz plots for income distribution in 1996 and 2007.  The data points are from IRS, and the theoretical curves represent \eref{lorenz_jump} with $f$ from \eref{f}.}
\label{fig:irslorenz}
\end{figure}

For the exponential distribution $P(r)=\exp(-r/T)/T$, it was shown in \cite{AAD01a} that the Lorenz curve is given by the formula $y=x+(1-x)\ln(1-x)$.  Notice that this formula is independent of $T$.  However, when the fat upper tail is present, this formula is modified as follows \cite{AAD03,Silva}
  \begin{equation}
  y = (1-f)[x + (1-x) \ln(1 -x)] + f\Theta(x-1).
  \label{lorenz_jump}
  \end{equation}
Here $\Theta(x-1)$ is the step function equal to 0 for $x<1$ and 1 for $x=1$.  The jump at $x=1$ is due to the fact that the fraction of population in the upper tail is very small, but their fraction $f$ of the total income is substantial.  

The data points in \fref{fig:irslorenz} show the Lorenz plots calculated from the IRS data for 1996 and 2007.   The solid lines in \fref{fig:irslorenz} are the theoretical Lorenz curves (\ref{lorenz_jump}) with the values of $f$ obtained from \eref{f}.  The theoretical curves agree well with the data.
The distance between the diagonal line and the Lorenz curve characterizes income inequality.  We observe in \fref{fig:irslorenz} that income inequality increased from 1995 to 2007, and this increase came exclusively from the growth of the upper tail, which pushed down the Lorenz curve for the exponential income distribution in the lower class.  

The standard way of characterizing inequality in the economic literature \cite{Kakwani-book} is the Gini coefficient $0\leq G\leq1$ defined as twice the area between the diagonal line and the Lorenz curve.  It was shown that $G=1/2$ for the exponential distribution \cite{AAD01a}, and 
  \begin{equation}
  G=\frac{1+f}{2}
  \label{Gini-f}
  \end{equation}
when taking into account the fraction $f$ going to the upper class on top of the exponential distribution \cite{Silva}.  The values of $G$ deduced from the IRS data
are given in \tref{table:IRS} and shown in panel (a) of \fref{fig:irsparam}
by the connected line, along with \eref{Gini-f} shown by open circles.  The increase of $G$ indicates that income inequality has been rising since 1983.  The agreement between the empirical values of $G$ and the formula \eref{Gini-f} in \fref{fig:irsparam} demonstrates that the increase in income inequality from the late 1990s comes from the upper tail growth relative to the lower class.

\section{Probability distribution of the global energy consumption}
\label{sec:energy}

\subsection{Introduction}

In the preceding sections, we studied monetary aspects of the economy and discussed probability distributions of money and income.  We found that significant inequality of money and income distributions can develop for statistical reasons.  Now we would like to discuss physical aspects of the economy.  Since the beginning of the industrial revolution several centuries ago, rapid technological development of the society has been based on consumption of fossil fuel, such as coal, oil, and gas, accumulated in the Earth for billions of years.  The whole discipline of thermodynamics was developed in physics to deal with this exploitation.  Now it is becoming exceedingly clear that these resources will be exhausted in the not-too-distant future.  Moreover, consumption of fossil fuel releases CO$_2$ to the atmosphere and affects the global climate.  These pressing global problems pose great technological and social challenges.

As shown below, energy consumption per capita by human population around the world has significant  variation.  This heterogeneity is a challenge and a complication for reaching a global consensus on how to deal with the energy problems.  Thus, it is important to understand and quantitatively characterize the global inequality of energy consumption.  In this section, we present such a study using the approach developed in the preceding sections of the paper.

\subsection{Energy consumption distribution as division of a limited resource}

Let us consider an ensemble of economic agents and characterize each agent $i$ by the energy consumption $\varepsilon_i$ per unit time.  Note that here $\varepsilon_i$ denotes not energy, but power, which is measured in kiloWatts (kW).  Similarly to \sref{sec:entropy}, we can discuss the probability distribution of energy consumption in the system and introduce the probability density $P(\varepsilon)$, such that $P(\varepsilon)\,d\varepsilon$ gives the probability to have energy consumption in the interval from $\varepsilon$ to $\varepsilon+d\varepsilon$.  Energy production, based on extraction of fossil fuel from the Earth, is physically limited.  So, energy production per unit time is a limited resource, which is divided for consumption among the global population.  As argued in \sref{sec:entropy}, it would be very improbable to divide this resource equally.  More likely, this resource would be divided according to the entropy maximization principle.  Following the same procedure as in \sref{sec:entropy}, with money $m$ replaced by energy consumption $\varepsilon$, we arrive at the conclusion that the probability distribution of $\varepsilon$ should follow the exponential law analogous to \eref{P(m)}
\begin{equation}
  P(\varepsilon) \propto e^{-\varepsilon/T},  \qquad T=\langle\varepsilon\rangle .
\label{P(e)}
\end{equation}
Here the ``temperature'' $T$ is the average energy consumption per capita.\footnote{To make it clear, this effective $T$ is not the temperature as it is known in physics.}

Now we would like to compare the theoretical conjecture \eref{P(e)} with the empirical data for energy consumption around the world.  For this purpose, it is convenient to introduce the cumulative distribution function 
  \begin{equation}
  C(\varepsilon)=\int_\varepsilon^\infty P(\varepsilon')\,d\varepsilon'.
  \label{C(e)}
  \end{equation}
Operationally, $C(\varepsilon)$ is the number of agents with the energy consumption above $\varepsilon$ divided by the total number of agents in the system.  If $P(\varepsilon)$ is an exponential function, then $C(\varepsilon)$ is also exponential.

\subsection{Empirical data analysis}

\begin{table}
\centering
\begin{tabular}{|l|c|r|r|r|r|r|r|}
\hline
Country       &  Label &   \multicolumn{3}{|c|}{Energy use (kW)} & \multicolumn{3}{|c|}{GDP/capita (k\$)} \\
\cline{3-8}
                    &        &  1990  &  2000   &  2005    &    1990   &    2000    &   2005   \\ \hline
Australia           &	AUS	 &	6.9	  &	 7.7	&	7.9    &	 18.9  &    20.9	&   36.3  \\
Bahrain             &	BHR	 &	13.0  &	 12.8	&	14.9   &    8.6   &    12.3   &	22.1  \\
Brazil              &	BRA	 &	1.2	  &	 1.4	&	1.5    &    3.1   &    3.7    &	4.7   \\
Canada              &	CAN	 &	10.0  &	 10.9	&	11.3   &    21.0  &    23.6   &	35.1  \\
China               &	CHN	 &	1.0	  &	 1.2	&	1.7    &     0.3   &     0.9	&	1.7   \\
Cuba                &	CUB	 & 	2.1	  &	 1.4	&	1.2    &	    &	 	&    \\
France              &	FRA	 &	5.3	  &	 5.8	&	6.0    &   21.9   &    22.4	&	35.0  \\
Germany             &	DEU	 &	6.0	  &	 5.6	&	5.6    &   21.6   &    23.1	&	33.7  \\
Iceland             &	ISL	 &	11.3  &	 15.3	&	16.3   &   24.5   &    30.9 	&	54.8  \\
India               &	IND	 &	0.5	  &	 0.6	&	0.6    &    0.4    &     0.4    &	0.7	   \\
Iran                &	IRN	 &	1.6	  &	 2.4	&	3.1    &    2.0   &    1.5	&	2.8   \\
Israel              &	ISR	 &	3.6	  &	 4.2	&	3.9    &   11.6   &    19.9	&	19.4  \\
Japan               &	JPN	 &	4.8	  &	 5.5	&	5.5    &   24.4   &    36.7   &	35.6  \\
Kenya               &	KEN	 &	0.7	  &	 0.6	&	0.7    &    0.4   &    0.4	 &	0.5	   \\
Kuwait              &	KWT	 &	5.3	  &	 12.2&	13.9   &    8.6   & 16.9	& 29.9  \\
Mexico              &	MEX	 &	2.0	  &	 2.0	&	2.3    &    3.1   &    5.8	& 7.4   \\
Netherlands Antilles &	ANT	 &	10.4  &	 10.2	&	11.9   &	    & 	&	    \\
Russia              &	RUS	 &	7.9	  &	 5.6	&	6.0    &   3.5    &    1.8	&	5.3   \\
Arab Emirates       &	ARE	 &	16.1  &	 14.7	&	15.2   &   18.0   &    21.7	&	31.6  \\
United Kingdom      &	GBR	 &	4.9	  &	 5.3	&	5.2    &   17.3   &    24.5	&	37.0  \\
United States       &	USA	 &	10.0  &	 10.8	&	10.4   &   22.5   &    34.3	&	41.3  \\
Qatar               &	QAT	 &	18.1  &	 25.6	&	26.5   &   15.8   &    28.8	&	53.3  \\ \hline
World average       &        &  2.2   &  2.2    &   2.3    &	4.2   &	5.2   &	7.0   \\ \hline
\end{tabular}
\caption{\label{table:WRI} Energy consumption per capita for the countries labeled in \fref{fig:eLinlin}, \fref{fig:eloglin}, and \fref{fig:eloglog}.  The units are converted from kilo tons of oil equivalent per year to kW (1 toe = $41.85\times10^9$ Joules).  The last three columns show GDP per capita from \cite{GDP}.}
\end{table}

We downloaded empirical data from the World Resources Institute (WRI) website \cite{WRI}.  The data on energy consumption is listed under the topic ``Energy and Resources''.  We downloaded the variable ``Total energy consumption'' \cite{EC}, which contains the annual energy consumption for various countries for the years 1990, 2000, and 2005 (only these years are available).  Population data is listed under the topic ``Population, Health and Human Well-being''.  We downloaded the variable ``Total population, both sexes'' \cite{TP}, which contains the total population of various countries for the same years.  From these two data files, we selected the countries for which both energy and population data are available.  Our final data files have 132 countries for 1990 and 135 countries for 2000 and 2005.  Then we divided the annual energy consumption in a given country by the population of this country to obtain the average energy consumption per capita $\varepsilon$.  The values of $\varepsilon$ are listed in \tref{table:WRI} for some countries.  A spreadsheet with our complete dataset is available for download as the supplementary online material of this paper.

\begin{figure}
\centering
\includegraphics[width=0.7\linewidth]{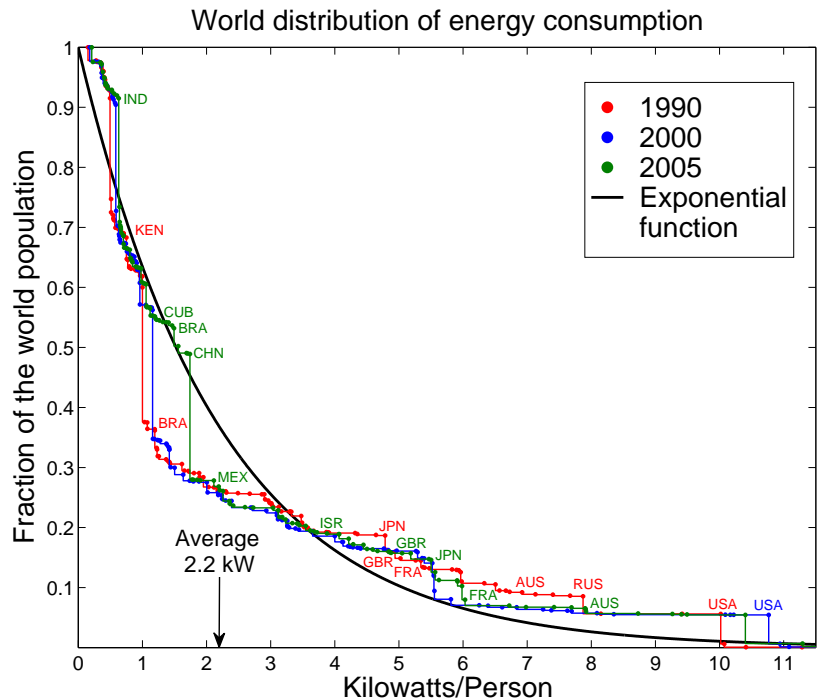}
\caption{Cumulative distribution functions of energy consumption per capita around the world for 1990, 2000, and 2005.  The solid curve is the exponential function. }
\label{fig:eLinlin}
\end{figure}

Then we proceeded to construct the cumulative probability distribution for $\varepsilon$.  First, we sorted the countries in the ascending order of their energy consumption per capita $\varepsilon_n$, so that $n=1$ corresponds to the country with the lowest consumption, and $n=L$ to the maximal consumption, where $L$ is the total number of countries.  We denoted the population of a country $n$ as $N_n$.  Then, the cumulative probability for a given $\varepsilon_n$ is 
  \begin{equation}
  C_{\rm e}(\varepsilon_n) = \frac{\sum_{k=n}^L N_k}{\sum_{k=1}^L N_k}.  
  \label{C(e_n)}  
  \end{equation}
Effectively, this construction assigns the same energy consumption $\varepsilon_n$ to all $N_n$ residents of the country $n$.  Of course, this is a very crude approximation, but it is the best we can do in the absence of more detailed data.  The empirically constructed function $C_{\rm e}(\varepsilon_n)$ is shown in \fref{fig:eLinlin} by different colors for the years 1990, 2000, and 2005.  \Tref{table:WRI} and \fref{fig:eLinlin} illustrate the great variation and inequality of energy consumption per capita around the world.  Let us focus on the data for 2005.  In USA, $\varepsilon$ is about 5 times greater than the global average; in China, $\varepsilon$ is close to the global average; and, in India, $\varepsilon$ is about 1/4 of the global average.  

By construction, $C_{\rm e}(\varepsilon_n)$ exhibits discontinuities at each $\varepsilon_n$ because of the approximation used in our procedure.  Given the relatively small number of data points ($L=135$) and discontinuities of the plot, it is not practical to do a quantitative fit of the data.  Nevertheless, the empirically constructed function $C_{\rm e}(\varepsilon)$ can be compared with the theoretical function $C_{\rm t}(\varepsilon)=\exp(-\varepsilon/T)$, which is shown by the solid line in \fref{fig:eLinlin}.  Here the temperature $T=2.2$~kW is the average global energy consumption per capita, obtained by dividing the total energy consumption of all countries by their total population.  This value is indicated by the arrow in \fref{fig:eLinlin}.  (For comparison, the physiological energy consumption at rest by a female of the weight 53 kg is 63 W \cite{Defilla-2007}.)   The exponential function does not fit the data perfectly, but it captures the main features reasonably well, given the crudeness of the data.  The agreement is remarkable, given that the solid line is not a fit, but a plot of a function with one parameter $T$ fixed by the global average.
  
\begin{figure}
\centering
\includegraphics[width=0.65\linewidth]{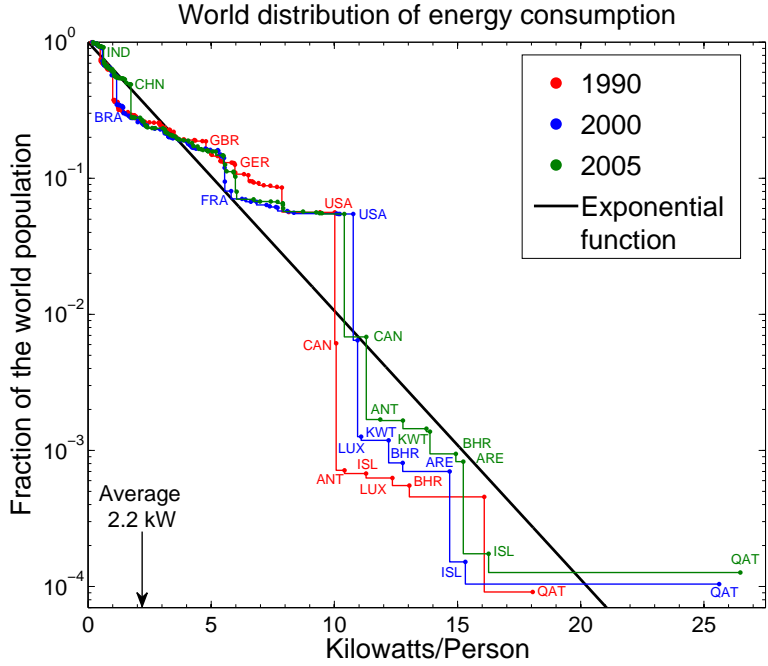}
\caption{The same data as in \fref{fig:eLinlin}, but plotted in the log-linear scale. }
\label{fig:eloglin}
\end{figure}

In order to make an additional visual comparison between the theory and the data, the functions $C_{\rm e}(\varepsilon_n)$ and $C_{\rm t}(\varepsilon)$ are plotted in \fref{fig:eloglin} in the log-linear scale and in \fref{fig:eloglog} in the log-log scale.  In \fref{fig:eloglin}, we see that the empirical data points oscillate around the theoretical exponential function shown by the straight line.  The data jumps for high $\varepsilon$ are unnaturally magnified in the logarithmic scale.  \Fref{fig:eloglog} demonstrates that the empirical data points do not fall on a straight line in the log-log scale, so the energy consumption per capita is not described by a power law.  Indeed, energy production and consumption are physically limited and have the characteristic average scale $T$, so a scale-free power-law distribution would not be expected here.

We have also constructed the plots for CO$_2$ emission per capita using the data from WRI \cite{WRI}.  They look essentially the same as the plots for energy consumption per capita, in agreement with findings by other authors \cite{MacKay}, because most of energy in the world is currently generated from fossil fuel.

\subsection{The effect of globalization on the inequality of energy consumption}

\begin{figure}
\centering
\includegraphics[width=0.65\linewidth]{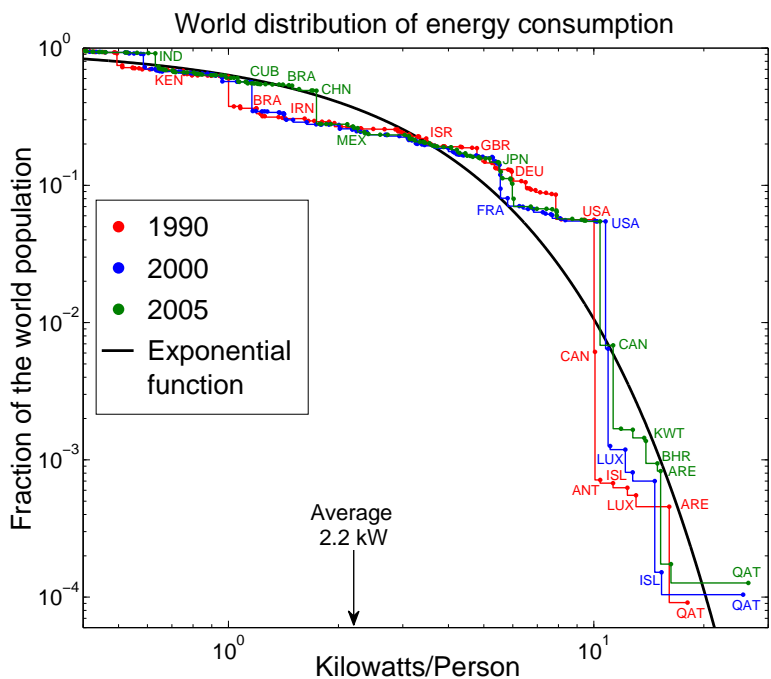}
\caption{The same data as in \fref{fig:eLinlin}, but plotted in the log-log scale.}
\label{fig:eloglog}
\end{figure}

\Fref{fig:eLinlin}, \fref{fig:eloglin}, and \fref{fig:eloglog} give different visual representations of $C(\varepsilon)$, but have certain shortcomings.  \Fref{fig:eLinlin} emphasizes the low end of the data, whereas \fref{fig:eloglin} and \fref{fig:eloglog} emphasize the high end.  All figures suffer from discontinuities. 

A smoother visualization can be achieved in the Lorenz plot for energy consumption per capita.  As in \eref{Lorenz}, the empirical Lorenz curve is constructed parametrically
  \begin{equation}
  x(\varepsilon_n) = \frac{\sum_{k=1}^n N_k}{\sum_{k=1}^L N_k}, \qquad
  y(\varepsilon_n) = \frac{\sum_{k=1}^n \varepsilon_k N_k}{\sum_{k=1}^L \varepsilon_k N_k}.
  \label{Lorenz-e}  
  \end{equation}
The horizontal coordinate $x(\varepsilon_n)$ gives the fraction of global population with energy consumption per capita below $\varepsilon$, and $y(\varepsilon_n)$ gives the total energy consumption of this population as a fraction of the global consumption.  When $n$ runs from 1 to $L$, we obtain a set of points in the $(x,y)$ plane representing the Lorenz plot.

\begin{figure}
\centering
\includegraphics[width=0.65\linewidth]{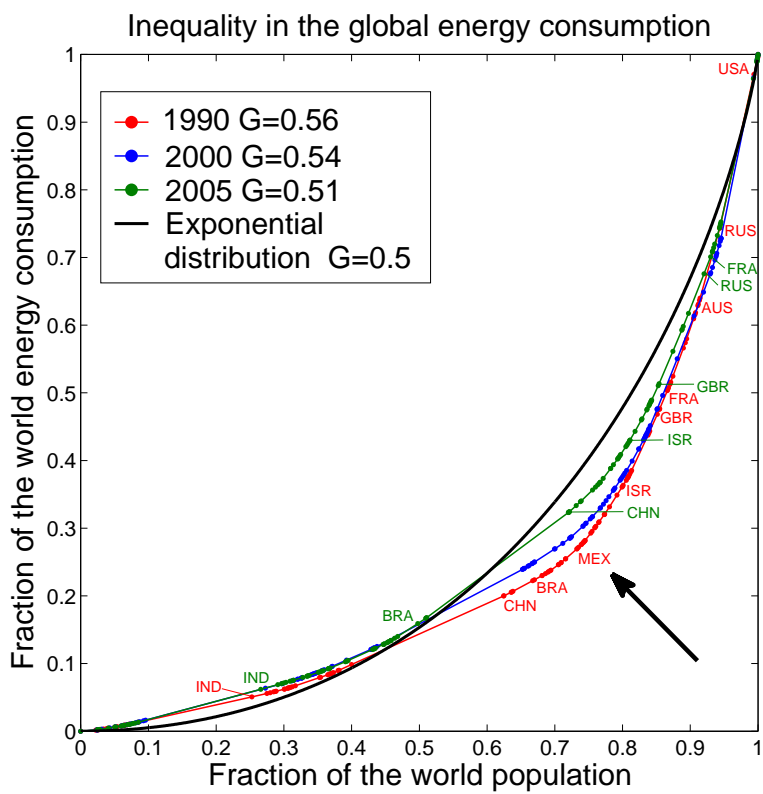}
\caption{Lorenz plots of energy consumption per capita around the world in 1990, 2000, and 2005, compared with the Lorenz curve for the exponential distribution.}
\label{fig:elorenz}
\end{figure}

The empirically constructed Lorenz plots for 1990, 2000, and 2005 are shown in \fref{fig:elorenz} using different colors.  By construction, the Lorenz plots are continuous without jumps, although the slope (the derivative) of the $y(x)$ curve is discontinuous.  Another advantage of the Lorenz plot is that it emphasizes the data where most of the population is, i.e., the range from the bottom 5\% to the top 95\% of the population sorted according to their energy consumption per capita.

The black solid line shows the theoretical Lorenz curve $y=x+(1-x)\ln(1-x)$ for the exponential distribution.  We observe that, in the first approximation, the theoretical curve captures the data reasonably well, especially given that the curve has no fitting parameters at all.  Upon a closer examination, we notice a systematic historical evolution of the empirical curves.  From 1990 to 2005, the data points moved closer to the diagonal, which indicates that global inequality of energy consumption decreased.  This is confirmed by the decrease of the calculated Gini coefficient $G$, which is listed in \fref{fig:elorenz}.

On the Lorenz plot for 1990, we notice a kink or a knee indicated by the arrow, where the slope of the curve changes appreciably.  This point represents the boundary between developed and developing countries.  Indeed, below this point we find Mexico, Brazil, China, and India, whereas above this point we find Britain, France, Australia, Russia, and USA.  The conclusion is that the difference between developed and developing countries lies in the degree of energy consumption and utilization.  This criterion provides a physical measure for such a distinction, as opposed to more ephemeral monetary measures, such as dollar income per capita.

Comparing the Lorenz plots for 2000 and 2005 with the plot for 1990, we observe that the kink in the plots is progressively smoothed out.  It means that the gap in energy consumption per capita between developed and developing countries is shrinking.  We attribute this result to rapid globalization of the world economy in the last 20 years.  Nevertheless, the distribution of energy consumption per capita around the world still remains highly unequal.  We observe in \fref{fig:elorenz} that  the Lorenz plot has moved closer to the solid curve representing the exponential distribution.  Based on the general arguments about partitioning of a limited resource, we expect that the  result of a well-mixed globalized world economy would not be an equal energy consumption, but the exponential distribution.  Thus, it is not likely that the energy consumption inequality will be eliminated in the foreseeable future.

It is generally known that energy consumption per capita and GDP per capita are positively correlated, and energy consumption is the physical basis for economic prosperity \cite{MacKay}.  Brown \textit{et al.}\ \cite{Brown-2010} found a power-law relation $\varepsilon\propto ({\rm GDP/capita})^{0.76}$ between these two variables by analyzing the data for different countries around the world (see figure 3A in \cite{Brown-2010}).   The last three columns in \Tref{table:WRI} show the data for GDP per capita \cite{GDP}.  Although this variable is generally correlated with the energy consumption per capita, the monetary and the physical measures are not always well aligned.  The movement of sustainable economics \cite{Daly-1996} criticized GDP as a useful measure of economic prosperity.

\section{Conclusions}

In this paper, we study probability distributions of money, income, and energy consumption per capita for ensembles of economic agents.  Following the principle of entropy maximization for partitioning of a limited resource among many agents, we find exponential distributions for the investigated variables.  Using an analogy with thermodynamics, we discuss trade deficit and immigration between two countries with different money temperatures.  Considering a cycle similar to a thermal engine, we discuss how a monetary profit can be  extracted in the presence of non-equilibrium due to a temperature difference.  

Then we study a Fokker-Planck equation for income diffusion with additive and multiplicative components.  The resulting probability distribution of income interpolates between the exponential function (Boltzmann-Gibbs) at the low end and the power law (Pareto) at the high end.  This function agrees well with the empirical income distribution data in USA obtained from the Internal Revenue Service.  While the exponential distribution in the lower class remains stable in time, the income fraction $f$ going to the upper tail expands dramatically during speculative bubbles and shrinks when the bubbles burst.  Overall, income inequality in USA has increased significantly from 1983 to 2007, so that now $f$ exceeds 20\% of the total income in the system.  We also discuss reasons why the Pareto exponent tends to have the value about $\alpha=2$ in the steady state in the absence of bubbles.

Finally, we analyze the probability distribution of energy consumption per capita around the world using the data from the World Resources Institute.  We find that the distribution is reasonably described by the exponential function with the average global consumption as the effective temperature.  A closer examination finds a gap in energy consumption between developed and developing countries, which tends to shrink as time progresses.  We attribute this effect to globalization of the world economy.  The inequality of energy consumption decreased from 1990 to 2005, while the corresponding Lorenz plot moved closer to the exponential distribution.

In conclusion, we observe that statistical problems of different nature have common mathematical description and exhibit similar and universal patterns of inequality.  Thus, statistical approach gives an insight into the persistent and ubiquitous nature of inequality in the world around us.  The approach presented here can be also applied to other statistical problems.

\ack

The authors are grateful to Doyne Farmer for hospitality at the Santa Fe Institute, where a part of this work was performed, and to the participants of the working group ``Universal Diversity Patterns Across the Sciences'' at the Santa Fe Institute in February 2009.  Victor Yakovenko also thanks Ted Jacobson for a stimulating discussion.

\section*{References}


\begin{thebibliography}{10}

\bibitem{RMP} Yakovenko V M and Rosser J B 2009 ``Colloquium: Statistical mechanics of money, wealth, and income'' \RMP {\bf 81} 1703

\bibitem{Wannier} Wannier G H 1987 \textit{Statistical Physics} (Dover, New York)

\bibitem{AAD00} Dr\u{a}gulescu A A and Yakovenko V M 2000 ``Statistical mechanics of money'' \emph{Eur. Phys. J. B} \textbf{17} 723

\bibitem{Mimkes-2000} Mimkes J 2000 ``Society as a many-particle system'' \textit{Journal of Thermal Analysis and Calorimetry} {\bf 60} 1055

\nonum Mimkes J and Willis G 2005 ``Lagrange principle of wealth distribution''  in \textit{Econophysics of Wealth Distributions} ed A Chatterjee, S Yarlagadda and B K Chakrabarti (Springer, Milan) pp 61--69

\bibitem{PhysNews} Sinha S and Chakrabarti B K 2009 ``Towards a physics of economics'' \textit{Physics News (Bulletin of Indian Physics Association)} v.39 No.2 (April) pp.33-46, available at \url{http://www.imsc.res.in/~sitabhra/publication.html}

\bibitem{CurrentScience} Chatterjee A, Sinha S and Chakrabarti B K 2007 ``Economic inequality: Is it natural?'' \emph{Current Science} \textbf{92} 1383

\bibitem{Schroeder} Schroeder D V 2000 \textit{An Introduction to Thermal Physics} (Addison Wesley Longman, New York) p 89

\nonum Yoshioka D 2007 \textit{Statistical Physics: An Introduction} (Springer, Berlin) p 25

\nonum Saha M and Srivastava B N 1931 \textit{A Treatise on Heat} (Indian Press, Allahabad) p 105 

\bibitem{Mirowski-book} Mirowski P 1989
  \textit{More Heat than Light: Economics as Social Physics, Physics as Nature's Economics} (Cambridge University Press, Cambridge)

\bibitem{Mimkes-2005} Mimkes J and Aruka Y 2005 ``Carnot process of wealth distribution'' in \textit{Econophysics of Wealth Distributions} ed A Chatterjee, S Yarlagadda and B K Chakrabarti (Springer, Milan) pp 70--78

\nonum Mimkes J 2006 ``A thermodynamic formulation of economics'' in \textit{Econophysics and Sociophysics: Trends and Perspectives} ed B K Chakrabarti, A Chakraborti and A Chatterjee (Wiley-VCH, Berlin) pp 1--33

\bibitem{Mimkes-2010} Mimkes J 2010 ``Stokes integral of economic growth: Calculus and the Solow model'' \textit{Physica A} {\bf 389} 1665

\bibitem{Vespignani} Serrano M A, Bogu\~n\'a M and Vespignani A 2007 ``Patterns of dominant flows in the world trade web'' \emph{Journal of Economic Interaction and Coordination} \textbf{2} 111

\bibitem{ITN} Bhattacharya K, Mukherjee G, Saram\"aki J, Kaski K and Manna S S 2008 ``The International Trade Network: weighted network analysis and modelling'' \emph{J. Stat. Mech.} P02002

\bibitem{AAD01a} Dr\u{a}gulescu A A and Yakovenko V M 2001 ``Evidence for the exponential distribution of income in the USA'' \emph{Eur. Phys. J. B} \textbf{20} 585

\bibitem{AAD01b} Dr\u{a}gulescu A A and Yakovenko V M 2001 ``Exponential and power-law probability distributions of wealth and income in the United Kingdom and the United States'' \emph{Physica} A \textbf{299} 213

\bibitem{AAD03} Dr\u{a}gulescu A A and Yakovenko V M 2003 ``Statistical mechanics of money, income, and wealth: a short survey'' in \textit{Modeling of Complex Systems: Seventh Granada Lectures} ed P L Garrido and J Marro, AIP Conf. Proc. \textbf{661} 180 (AIP, New York)

\bibitem{Pareto-book} Pareto V 1897 \textit{Cours
  d'\'Economie Politique} (L'Universit\'e de Lausanne)

\bibitem{Silva} Silva A C and Yakovenko V M 2005 ``Temporal evolution of the `thermal' and `superthermal' income classes in the USA during 1983-2001'' \emph{Europhys. Lett.} \textbf{69} 304

\bibitem{Kinetics} Lifshitz E M and Pitaevskii L P 1981 \emph{Physical Kinetics} (Pergamon Press, Oxford)

\bibitem{Gibrat-1931} Gibrat R 1931 \textit{Les In\'egalit\'es Economiques} (Sirely, Paris)

\bibitem{Aoki-2003} Fujiwara Y, Souma W, Aoyama H, Kaizoji T and Aoki M 2003 ``Growth and fluctuations of personal income'' \emph{Physica A} \textbf{321} 598
  
\nonum Aoyama H, Souma W and Fujiwara Y 2003 ``Growth and fluctuations of personal and company's income'' \emph{Physica A} \textbf{324} 352

\bibitem{Fiaschi-Marsili} Fiaschi D and Marsili M 2009 ``Economic interactions and the distribution of wealth'' arXiv:0906.1512

\bibitem{IRS-data} Personal income data from the Statistics of Income division of IRS \url{http://www.irs.gov/taxstats/indtaxstats/article/0,,id=134951,00.html}

\bibitem{Dent-1993} Dent H S 1993 \textit{The Great Boom Ahead} (Hyperion, New York)

\bibitem{Dent-2009} Dent H S 2009 \textit{The Great Depression Ahead} (Free Press, New York)

\bibitem{PWA} Anderson P W A 1997 ``Some thoughts about distribution in economics'' in \textit{The Economy as an Evolving Complex System II} ed W B Arthur, S N Durlauf and D A Lane (Addison-Wesley, Reading) pp 565--566

\bibitem{earthquake} Bhattacharyya P, Chatterjee A and Chakrabarti B K 2007 ``A common mode of origin of power laws in models of market and earthquake'' \emph{Physica A} \textbf{381} 377

\bibitem{Kakwani-book} Kakwani N 1980
  \textit{Income Inequality and Poverty} (Oxford University Press,
  Oxford)

\bibitem{WRI} World Resources Institute data \url{http://earthtrends.wri.org/}

\bibitem{EC} Link to ``Total energy consumption'' 
\url{http://earthtrends.wri.org/searchable_db/index.php?theme=6&variable_ID=267&action=select_countries}

\bibitem{TP} Link to ``Total population, both sexes'' 
\url{http://earthtrends.wri.org/searchable_db/index.php?theme=4&variable_ID=363&action=select_countries}

\bibitem{Defilla-2007} Defilla S 2007 ``A natural value unit -- Econophysics as arbiter between finance and economics'' \textit{Physica A} {\bf 382} 42

\bibitem{MacKay} MacKay D J C 2009 \textit{Sustainable Energy -- without the hot air} (UIT, Cambridge, England)

\bibitem{Brown-2010} Brown J H, Burnside W R, Davidson A D, DeLong J P, Dunn W C, Hamilton M J, Nekola J C, Okie J G, Mercado-Silva N, Woodrufff W H and Zuo W 2010 ``Energetic limits to economic growth'' \emph{Proceedings of the National Academy of Sciences of the USA}, in press

\bibitem{GDP} Link to the GDP data \url{http://earthtrends.wri.org/searchable_db/index.php?theme=5&variable_ID=224&action=select_countries}				

\bibitem{Daly-1996} Daly H 1996 \emph{Beyond Growth: The Economics of Sustainable Development} (Beacon Press, Boston)

\end{thebibliography}
\end{document}